\documentstyle[12pt,fleqn]{article}
\author{ L.~Didukh\\
{\small \it Ternopil State Technical University, Department of Physics}\\ 
{\small \it 56 Rus'ka Str., Ternopil UA--282001, Ukraine}\\
{\small \it Tel.:+380352251946, Fax: +380352254983}\\
{\small \it E-mail: didukh@tu.edu.te.ua}}
\date{}
\title{Ferromagnetic ordering in a generalized Hubbard model:
weak intra-atomic interaction limit}
\begin{document}
\maketitle

\begin{abstract}
In the present work ferromagnetic ordering in the Hubbard 
model generalized by taking into account the inter-atomic exchange 
interaction and correlated hopping 
in partially filled narrow band is considered.
Expressions for the magnetization and Curie temperature as functions 
of model parameters and band filling are obtained in the
case of weak intra-atomic Coulomb interaction. 
Condition of ferromagnetic state realization is found. 
The obtained results indicate the important role of correlated hopping.
pacs{71.10.Fd, 71.30.+h, 71.27.+a}
\end{abstract}

\section{Introduction}

In spite of great attention payed to investigation of ferromagnetism 
in narrow energy bands~\cite{1_2}-\cite{1_54}, where the same system 
of electrons is responsible both 
for conductivity and for magnetic ordering (the transition $3d$-metals, 
ferromagnetic sulphides, oxides and selenides with metallic type of 
conductivity) the problem still remains open~\cite{amad,hirsch}.  

There are some experimental results for the disulphides of transition metals
which can be interpreted on the basis of itinerant mechanism of 
ferromagnetism. In these compounds the dependence of Curie 
temperature on the number of magnetic moments is non-typical 
from point of view of exchange mechanism of feromagnetism. 
In $Fe_{x}Co_{1-x}S_{2}$ ~\cite{jar} where electron concentration $n$
changes from 0 to 1 in doubly degenerate $e_g$ -subband
the Curie temperature increases with decrease of electron concentration
at $0.75<n<1$.

Although the consistent theory of ferromagnetic ordering in  transition
metal compounds can be constructed only in the model including the orbital 
degeneracy of the band, the qualitative character of observed properties 
can be interpreted in the framework of generalized model of non-degenerate
band~\cite{did4}.

\section{The Hamiltonian}
\setcounter{equation}{0}
The simplest model for description of the magnetic properties of narrow-band
materials is the Hubbard model~\cite{1_1}, but this model contains only 
diagonal matrix elements of Coulomb interaction in site representation, 
that not give rise to metallic ferromagnetism except in special situations
such as a single hole in a half-filled band~\cite{nag} or a special lattice 
geometry~\cite{1_54}.  The importance of non-diagonal  matrix elements of 
Coulomb interaction was pointed out in papers~\cite{1_12,amad}. 

We write the Hamiltonian of system of $s$-electrons in the 
following form~\cite{cmp,jps1}
\begin{eqnarray} \label{Ham}
H=&-&\mu \sum_{i\sigma}a_{i\sigma}^{+}a_{i\sigma}+
{\sum_{ij\sigma}}' t_{ij}(n)a_{i\sigma}^{+}a_{j\sigma}+
+{\sum_{ij\sigma}}'\left(T_2(ij)a_{i\sigma}^{+}a_{j\sigma}n_{i\bar{\sigma}}
+h.c.\right)
\\ \nonumber
&+&U\sum_{i}n_{i\uparrow}n_{i\downarrow}
+{1\over 2}{\sum_{ij\sigma \sigma^{'}}}'J a_{i\sigma}^{+}
a_{j\sigma{'}}^{+}a_{i\sigma{'}}a_{j\sigma},
\nonumber
\end{eqnarray}
where $a_{i\sigma}^{+}$, $a_{i\sigma}$ --  creation
and destruction operators of electron on site $i$, 
$\sigma =\uparrow, \downarrow$, $n_{i\sigma}=a_{i\sigma}^{+}a_{i\sigma}$,
$n=\langle n_{i\uparrow}+n_{i\downarrow}\rangle$,
$\mu$ -- chemical potential, $t_{ij}(n)=t(ij)+T_1(ij)$
is the effective hopping integral between nearest neighbours,
$t_{ij}$ is the hopping integral of an electron 
from site $j$ to site $i$, $T_1(ij)$ and $T_2(ij)$ are the parameters 
of correlated hopping~\cite{jps1} 
of electron, $U$ is the intraatomic Coulomb repulsion and $J$
is the exchange integral for the nearest neighbours.   
The prime by second sum in 
Eq.~(\ref{Ham}) signifies that $i\neq{j}$. 

The peculiarities of the model described by the Hamiltonian~(\ref{Ham}) 
is taking into consideration the influence of sites occupation on the 
electron hoppings (correlated hopping), and the exchange integral.
The correlated hopping, firstly, renormalize the initial hopping 
integral (it becomes concentration- and spin-dependent) and, secondly,
lead to an independent on quasiimpulse shift of the band center, dependent on
magnetic ordering. Taking into account the quantity of second order $J$ 
is on principle necessary to describe ferromagnetism in this 
model~\cite{1_12,amad,camp}.
In this paper we do not take into account the 
inter-atomic Coulomb interaction, which is important in the charge ordering 
state (this ordering type go beyond this article). 

To characterize the value of correlated hopping we introduce dimensionless
parameters $\tau_1=\frac{T_1(ij)}{|t_{ij}|}$, 
$\tau_2=\frac{T_2(ij)}{|t_{ij}|}$.

\section{Weak intra-atomic interaction}

To simplify the consideration we use model Hamiltonian~(\ref{Ham}). If 
intra-atomic Coulomb interaction is weak ($U<|t_{ij}(n)|$) 
then we can take into account 
the electron-electron interaction in the Hartree-Fock approximation:
\begin{eqnarray}\label{hfa}
&&n_{i\uparrow}n_{i\downarrow}=n_{\uparrow}n_{i\downarrow}+
n_{\downarrow}n_{i\uparrow},
\\
\nonumber
&&a_{i\sigma}^{+}n_{i\bar{\sigma}}a_{j\sigma}=n_{\bar{\sigma}}
a_{i\sigma}^{+}a_{j\sigma}+
\langle a_{i\sigma}^{+}a_{j\sigma}\rangle n_{i\bar{\sigma}},
\end{eqnarray}
where the average values $\langle n_{i\sigma}\rangle=n_{\sigma}$ 
are independant
of site number (we suppose that distributions  of electron charge and 
magnetic momentum are homogenous). Taking into account~(\ref{hfa}) we can write 
Hamiltonian~(\ref{Ham}) in the following form:
\begin{eqnarray}\label{ham1}
&& 
H={\sum}'_{ij\sigma}\epsilon_\sigma(ij)a_{i\sigma}^{+}a_{j\sigma},
\end{eqnarray}
where
\begin{eqnarray}
&& \epsilon_\sigma(ij)=-\mu+\beta_\sigma+n_{\bar{\sigma}}U + zn_{\sigma}J +
t_{ij}(n\sigma);
\end{eqnarray}

\begin{eqnarray}
&& \beta_\sigma={2 \over N}\sum_{ij} T_2(ij) \langle 
a_{i\bar{\sigma}}^{+}a_{j\bar{\sigma}}\rangle,
\end{eqnarray}

\begin{eqnarray}
&& t_{ij}(n\sigma)=t_{ij}(n)+2n_{\bar{\sigma}}T_2(ij).
\end{eqnarray}
The dependences of effective hopping integral on electron concentration 
and magnetization, a being of the spin-dependent displasement of band
center are the essential distinction of single-particle energy spectrum
in the model described by Hamiltonian~(\ref{ham1}) from the spectrum in the 
Hubbard model for weak interaction. An use of~(\ref{ham1}) allows, in particular,
to explain the peculiarities of dependence of binding energy on 
atomic number in transition metals and also essentially modifies
theory of ferromagnetism in a collective electron model.

\section{Ferromagnetic ordering in weak intra-atomic interaction limit}

After the transition to Fourier representation we obtain for the Green 
function 
\begin{eqnarray}
&& \langle\langle a_{p\sigma}|a^{+}_{p'\sigma}\rangle\rangle_{\bf k}
=\frac{1}{2\pi}\frac{1}{E-\epsilon_\sigma(\bf k)}
\end{eqnarray}
where the energy spectrum is
\begin{eqnarray}\label{spectr}
&& \epsilon_\sigma({\bf k})=-\mu+\beta_\sigma+n_{\bar{\sigma}}U-zn_{\sigma}J+
t(n\sigma)
\gamma({\bf k});
\end{eqnarray}
here the spin-dependent shift of the band center is
\begin{eqnarray}
&& \beta_\sigma={2 \over N}\sum_{ij} T_2(ij) \langle 
a_{i\bar{\sigma}}^{+}a_{j\bar{\sigma}}\rangle,
\end{eqnarray}
$\gamma({\bf k})=
\sum\limits_{{\bf R}}e^{i{\bf kR}}$,
the spin and concentration dependent hopping integral is
\begin{eqnarray}
&& t(n\sigma)=t(n)+2n_{\bar{\sigma}}T_2.
\end{eqnarray}

The concentration of electrons with spin $\sigma$ is 
\begin{eqnarray}
n_{\sigma}=\int\limits_{-\infty}^{+\infty}\rho(\epsilon)f(\epsilon)d\epsilon.
\end{eqnarray}
Here $\rho(\epsilon)$ is the density of states, $f(\epsilon)$ is the Fermi 
distribution function. Let us assume the rectangular density of states:
\begin{eqnarray}
\rho(\epsilon)=\frac{1}{N}\sum_{\bf k}\delta(\epsilon-\epsilon({\bf k}))=
\frac{1}{2w}\theta(\epsilon^{2}-w^{2})
\end{eqnarray}

 Then in the case of zero temperature we obtain:
\begin{eqnarray}
n_{\sigma}=\int\limits_{-\infty}^{+\infty}\rho(\epsilon)\theta(-E_{\sigma}(
\epsilon))d\epsilon=\frac{1}{2w}\int\limits_{-w}^{w}\theta(-E_{\sigma}(\epsilon))
d\epsilon=\frac{1}{2w}\int\limits_{-w}^{\epsilon_{\sigma}}d\epsilon,
\end{eqnarray}
where 
$\epsilon_{\sigma}=w(2n_{\sigma}-1)$.
The value
$\epsilon_{\sigma}$ is the solution of the equation $E_{\sigma}(\epsilon)=0$
from which we obtain
$\epsilon_{\sigma}=\frac{\mu_{\sigma}}{\alpha_{\sigma}}$,
where
$\mu_{\sigma}=\mu-\beta_{\sigma}+zn_{\sigma}J-n_{\bar\sigma}U$
and
$\alpha_{\sigma}=1-2\tau_2 n_{\bar\sigma}$.

The system parameters are related by the equation
\begin{eqnarray}
\label{steq}
m(zJ+u)+\beta_{\downarrow}-\beta_{\uparrow}=2m(1-\tau_2).
\end{eqnarray}

The shift of band center is obtained from
\begin{eqnarray}
\beta_{\sigma}=\frac{2}{N}\sum_{ij}T_2(ij)\langle a^{+}_{i \bar\sigma}
a_{j \bar\sigma} \rangle=-\frac{\tau_2}{2w}\int\limits_{-w}^{\epsilon_
{\sigma}}\epsilon d\epsilon=-\tau_2 wn_{\bar\sigma}(n_{\bar\sigma}-1).
\end{eqnarray}

One can see that
\begin{eqnarray}
\label{beta}
\beta_{\downarrow}-\beta_{\uparrow}=2\tau_2 mw(1-n).
\end{eqnarray}

From the equation~(\ref{steq}) one can see that in less than 
half-filled band correlated 
hopping leads to the stabilisation of ferromagnetism as well as 
inter-atomic exchange interaction and intra-atomic Coulomb interaction;
the larger is electron concentration $n$ the smaller is the  
critical value of exchange integral for occurence of ferromagnetism.
These our results are in accordance with the results of 
paper~\cite{amad}. 

From (\ref{steq}) and (\ref{beta}) we obtain the condition of feromagnetic
ordering realization:
\begin{eqnarray}
\frac{zJ+U}{2w}+\tau_2(2-n)>1.
\end{eqnarray}
Here we use the notation
\begin{eqnarray}
w=w(n)=z|t_{0}|(1-\tau_{1}n)
\end{eqnarray}
where $t_{0}$ is the band hopping integral.

In the case of non-zero temperatures the equation for magnetization is 
\begin{eqnarray}
\label{magnet}
\exp\left({\frac{-mJ_{eff}}{\theta}}\right)=\frac{\sinh\left((1-n_{\uparrow})
\alpha_{\uparrow}w/ \theta\right)\sinh\left(n_{\downarrow}\alpha_{\downarrow}
w/ \theta\right)}
{\sinh((1-n_{\downarrow})\alpha_{\downarrow}w/ \theta)\sinh(n_{\uparrow}
\alpha_{\uparrow}w/ \theta)},
\end{eqnarray}
where the effective exchange integral is
\begin{eqnarray}
J_{eff}=zJ+U+2zT_2(1-n)
\end{eqnarray}

The expression for Curie temperature can be obtained from 
equation~(\ref{magnet}) at $m\rightarrow 0$.  If $\tau_1=\tau_2=0$ then
we obtain
\begin{eqnarray}
\theta_{C}=\frac{w_{0}}{2}{\rm arcth}^{-1}\left(\frac{2w_{0}}{zJ+U}\right)
\end{eqnarray}
If $\tau_{1}=\tau_{2}=1/2$ then
\begin{eqnarray} \label{Tc}
\theta_{C}=\frac{w_{0}n(2-n)(1-{n\over 2})}{4} 
{\rm arcth}^{-1}\left(\frac{w_{0}(1-{n\over 2})}
{zJ+U+(1-n)(1-{n\over 2})w_{0}}\right)
\end{eqnarray}
It is important that at some values of parameters the Curie temperature 
can increase with the decrease of the electron concentration.

\section{Discussion and Conclusions}
The analisys of obtained in this paper energy spectrum~(\ref{spectr})
shows that both intraatomic Coulomb repulsion $U$ and 
exchange integral $J$ favor spin polarization. Futhermore,
taking into account the correlated hopping leads to the 
spin-dependent shift of the band center and to the band narrowing,
what also give rise to ferromagnetism. These our results are in 
accordance with the results of paper~\cite{amad}. 

The peculiarities of the energy spectrum~(\ref{spectr}) 
lead to concentration dependence of the Curie temperature. 
In particular, the concentration dependent shift~(\ref{beta}) 
of the band centers at $n<1$ is positive, at $n>1$ is negative.
According to this fact the Curie temperature at decreasing
$n$ from 1 can increase and at $n>1$ can decrease.
 The obtained dependence qualitaively agrees with the
experimental data on Fe$_x$Co$_{1-x}$S$_2$ \cite{jar}.
The proposed approach can be extended to all values of $n$,
and the peculiarities of ferromagnetic ordering in Co$_x$Ni$_{1-x}$S$_2$ 
where this concentration  changes from 1 to 2 can be explained.

In conclusion, taking into consideration both correlated hopping and 
inter-atomic exchange interaction essentially modify the $s-$band model
and favours the ferromagnetic ordering.

\end{document}